\documentclass[lettersize,journal]{IEEEtran}
\usepackage{amsmath,amsfonts}
\usepackage{algorithmic}
\usepackage{array}
\usepackage{subfigure}
\usepackage{textcomp}
\usepackage{stfloats}
\usepackage{url}
\usepackage{verbatim}
\usepackage{graphicx}
\def\BibTeX{{\rm B\kern-.05em{\sc i\kern-.025em b}\kern-.08em
    T\kern-.1667em\lower.7ex\hbox{E}\kern-.125emX}}
\usepackage{balance}
\usepackage{adjustbox}
\usepackage{amssymb}
\usepackage{booktabs}
\usepackage{multirow}

\begin{document}
\title{Covariance-Aware Detection for Single-RF MIMO-OFDM Systems}

\author{Tianrui Qiao,~\IEEEmembership{Member,~IEEE,} 
	Jun Qian,~\IEEEmembership{Member,~IEEE,} 
	and Ross Murch,~\IEEEmembership{Fellow,~IEEE}
	\thanks{Manuscript received; This work was supported by the Hong Kong Research Grants Council for the General Research Fund (GRF) grant 16208124. (\emph{Corresponding author: Jun Qian.})}
	\thanks{The authors are with the Department of Electronic and Computer Engineering, The Hong Kong University of Science and Technology, Hong Kong, China (email: eetqiao@ust.hk; eejunqian@ust.hk; eermurch@ust.hk).}}

\markboth{IEEE Transactions on Vehicular Technology}{QIAO \MakeLowercase{\emph{et al.}}: Covariance-Aware Detection for Single-RF MIMO-OFDM Systems}

\maketitle

\begin{abstract}
We propose the use of a covariance-aware detection method for single-radio-frequency (RF) multiple-input multiple-output (MIMO) orthogonal frequency division multiplexing (OFDM) systems. This detection approach enhances bit error rate (BER) performance as compared to conventional detectors. The approach is motivated by the observation that when using reconfigurable antennas (such as electronically steerable parasitic array radiators (ESPARs)) to create a single-RF MIMO system, an additional model error arising from the reconfigurable antennas is introduced. These modeling errors produce colored noise and an irreducible BER (error floor) at high signal-to-noise ratios (SNRs). Simulation results, using ESPAR as an example, validate our error floor analysis and demonstrate that the use of the proposed enhanced detection method can effectively address the error floor and improve the BER at high transmit SNRs.  
\end{abstract}
\vspace{-0.1cm}

\vspace{-0.2cm}
\section{Introduction}
\IEEEPARstart{S}{ingle}-radio-frequency (RF) multiple-input multiple-output (MIMO) systems are a promising alternative to conventional MIMO for providing reduced complexity and improved energy efficiency, due to the reduction in the number of RF chains \cite{SingleRFIntro1,SingleRFIntro2}. Approaches to achieve single-RF MIMO include utilizing reconfigurable antennas \cite{ReconAntennaIntro1,ReconAntennaIntro2} such as load modulated arrays \cite{LM}, pixel antennas \cite{AntennaCoding1,AntennaCoding2}, and electronically steerable parasitic array radiators (ESPARs) \cite{ESPAR}.      

An ESPAR is a type of reconfigurable antenna that consists of an active antenna connected to the single-RF chain with passive parasitic antennas loaded with variable reactances surrounding it. The parasitic antennas are coupled to the active antenna through their close  proximity to the active element \cite{ESPAR}. Tuning the reactive loads changes the currents on the parasitic antennas, yielding different radiation patterns of the ESPAR \cite{ESPAR,FlatCurrent2Li,HAN_OFDM}. Therefore, an ESPAR can be thought of as a reconfigurable antenna and employed to transmit signals in MIMO systems, where the currents and the patterns of ESPAR are shaped according to the transmitted signals \cite{FlatCurrent2Li,HAN_OFDM}. Such ESPAR enabled MIMO systems have been considered in frequency-flat propagation environments \cite{FlatCurrent2Li,FlatCurrent1,ESPARbs1,Flatbs} and frequency-selective channels implementing orthogonal frequency division multiplexing (OFDM) schemes \cite{HAN_OFDM,OFDMCurrent}. 

When evaluating the symbol or the bit error rate (BER) performance in these single-RF MIMO systems employing ESPAR, an error floor has been observed \cite{FlatCurrent2Li,HAN_OFDM,FlatCurrent1}. This is because the currents on ESPAR do not perfectly match the ideal currents required by the signals to be transmitted in practice. This is due to non-ideal reactances which are not continuously variable for example. To alleviate the performance degradation caused by the error floor, additional operations at the transmitter have been proposed in frequency-flat channel scenarios. These include exciting the ESPAR with an optimized voltage source \cite{FlatCurrent2Li} or pre-processing the signals for transmission \cite{FlatCurrent1}. However, the observed error floor in single-RF MIMO systems has not been systematically explored in frequency-selective propagation environments. Furthermore, alleviation of the error floor has only been considered at the transmitter \cite{FlatCurrent2Li,FlatCurrent1} while the possibility for addressing the error floor at the receiver has not been exploited.   

To address these research gaps, we systematically investigate and analyze the error floor in single-RF MIMO-OFDM systems using the beamspace concept for frequency-selective channels, when ESPAR is employed at the transmitter as a reconfigurable antenna. To solve the error floor problem, we propose applying a covariance-aware detection approach for handling the colored noise in single-RF MIMO systems \cite{ML1,ML2}. The technique is robust to the model error introduced by using ESPAR to transmit signals. We also provide simulation results under various communication scenarios to validate our error floor analysis and demonstrate the effectiveness and generality of the proposed detection approach for low-complexity single-RF uplink transmission. 

\textit{Notation}: Bold lower and upper case letters denote vectors
and matrices, respectively. Upper case letters in calligraphy represent
sets. Letters not in bold font are scalars. $\mathbb{C}$ is the complex number set. $j=\sqrt{-1}$ is the imaginary unit. $\mathcal{CN}\left(\mu,\sigma^{2}\right)$ is a complex Gaussian distribution with mean $\mu$ and standard deviation $\sigma$. $\left|a\right|$ denotes the absolute value of a scalar $\mathit{a}$. $\mathrm{diag}\left(a_{1},\ldots,a_{N}\right)$ is a diagonal matrix with diagonal entries being $a_{1},\ldots,\mathit{a_{N}}$. $\lVert \mathbf{a} \rVert$ is the Euclidean norm of a vector $\mathbf{a}$. $\langle\mathbf{a},\mathbf{b}\rangle$ denotes the inner product of vector $\mathbf{a}$ and vector $\mathbf{b}$. $\mathbf{U}_N$ indicates an $N\times N$ identity matrix. $\mathbb{E}\{\cdot\}$, $(\cdot)^{\mathit{T}}$, $(\cdot)^{\mathit{H}}$, and $(\cdot)^{-1}$ represent the expectation, transpose, conjugate transpose, and inverse, respectively.

\vspace{-0.2cm}
\section{Problem Formulation}
\subsection{System Model}
\begin{figure}[t]
	\centering
	\includegraphics[scale=0.46]{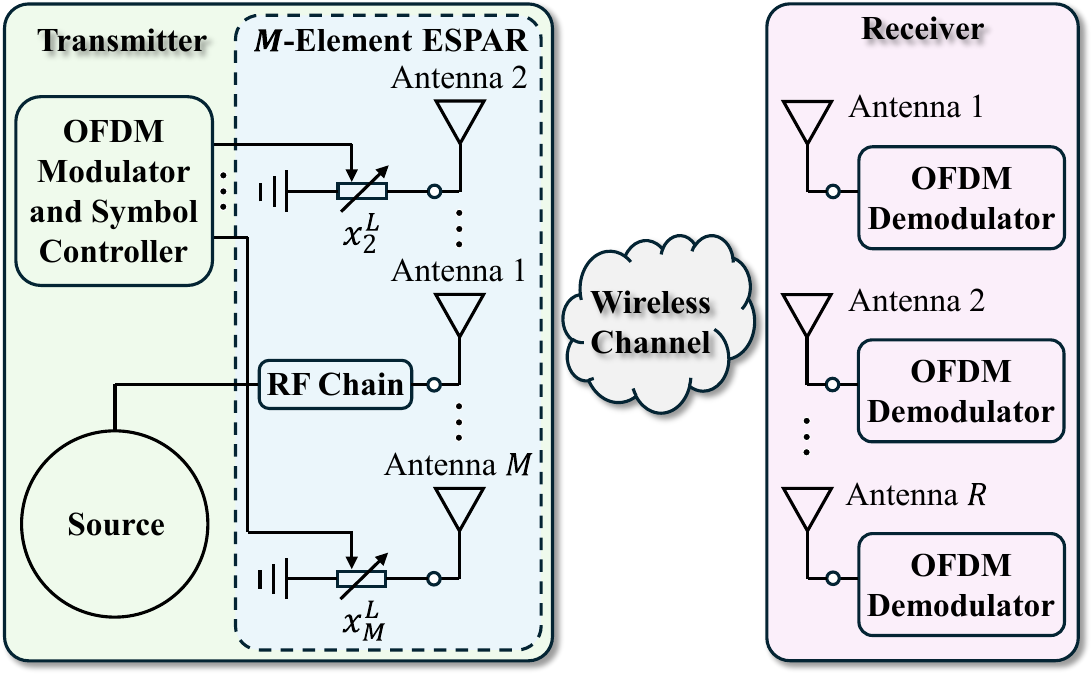}
	\vspace{-0.3cm}
	\caption{Schematic of the MIMO-OFDM system using a single-RF $M$-element ESPAR with $M-1$ variable reactive loads at the transmitter and $R$ conventional antennas at the receiver.}
	\vspace{-0.5cm}
	\label{Fig_System_Model}
\end{figure}

A MIMO-OFDM system using an $M$-element single-RF ESPAR at the transmitter is presented in Fig. \ref{Fig_System_Model}, where the active antenna 1 is connected to a single-RF source and the $M-1$ parasitic antennas are connected to variable reactive loads. It has been shown that an $M$-element ESPAR can excite $N\leq M$ orthogonal far-field radiation patterns \cite{PCDM}. They can also be thought of as a pattern basis set when the orthogonal patterns are normalized to unit radiated power. The normalized patterns are denoted as $\mathbf{E}_\mathrm{bs}=[\mathbf{e}_1^\mathrm{bs},\ldots,\mathbf{e}_N^\mathrm{bs}]\in\mathbb{C}^{2J\times N}$ satisfying $\mathbf{E}^H_\mathrm{bs}\mathbf{E}_\mathrm{bs}=\mathbf{U}_N$, where $\mathbf{e}_n^\mathrm{bs}\in\mathbb{C}^{2J\times1}$, $n\in\mathcal{N}=\{1,\ldots,N\}$, is the $n$th basis pattern containing $\theta$ and $\phi$ polarization components uniformly sampled over $J$ spatial angles. $N$ can also be regarded as the effective aerial degrees-of-freedom (EADoF) of the ESPAR and becomes an estimate for the number of independent streams of OFDM symbols that can be simultaneously transmitted \cite{PCDM,Sayeed,ESPARbs2DoF}.  

We denote $\mathbf{s}_k=[s_{k,1},\ldots,s_{k,N}]^T\in\mathbb{C}^{N\times1}$ as the transmitted signal in the frequency domain with $s_{k,n}$, $\forall n\in\mathcal{N}$, being related to the $n$th basis pattern at the $k$th subcarrier, $\forall k\in\mathcal{K}=\{1,\ldots,K\}$, where $K$ indicates the number of subcarriers used for transmission. We assume that the wireless propagation environment experiences $L$-tap frequency selective fading while the antennas at the transmitter and the receiver are frequency-flat. The cyclic prefix (CP) having length $v$ is assumed to be sufficiently long so that there is no inter-symbol interference. $R$ ideally isolated conventional antennas are used for receiving the signals. Leveraging the beamspace concept, by modulating the transmitted signals onto the orthogonal basis patterns of ESPAR, the frequency-domain MIMO-OFDM system can be formulated as \cite{AntennaCoding2,HAN_OFDM}
\begin{equation}
	\mathbf{y}_k = \mathbf{F}_{\mathrm{bs}}^T\mathbf{H}_{\mathrm{v},k}\mathbf{E}_\mathrm{bs}\mathbf{s}_k+\mathbf{n}_k = \mathbf{H}_{\mathrm{bs},k}\mathbf{s}_k + \mathbf{n}_k,\;\forall k\in\mathcal{K},
	\label{system_model_freq}
\end{equation}
where $\mathbf{F}_\mathrm{bs}=[\mathbf{f}_1^\mathrm{bs},\ldots,\mathbf{f}_R^\mathrm{bs}]\in\mathbb{C}^{2J\times R}$ collects the $R$ receiver orthogonal basis radiation patterns $\mathbf{f}_r^\mathrm{bs}\in\mathbb{C}^{2J\times1}$ containing the $\theta$ and $\phi$  polarization components sampled over $J$ angles and satisfying $\mathbf{F}^H_\mathrm{bs}\mathbf{F}_\mathrm{bs}=\mathbf{U}_R$. $\mathbf{y}_k\in\mathbb{C}^{R\times1}$ denotes the received signal at the $k$th subcarrier and $\mathbf{n}_k\in\mathbb{C}^{R\times1}$ is the spectrally-uncorrelated additive white Gaussian noise (AWGN) at the $k$th subcarrier, which follows independently and identically distributed (i.i.d.) $\mathcal{CN}(\mathbf{0},N_0\mathbf{U}_R)$ with $N_0$ referring to the noise power. $\mathbf{H}_{\mathrm{v},k}\in\mathbb{C}^{2J\times2J}$ is the virtual beamspace channel matrix at the $k$th subcarrier, collecting the channel gain from each angle of departure to each angle of arrival. The frequency-domain virtual channel $\mathbf{H}_{\mathrm{v},k}$ can be obtained from the corresponding time-domain $L$-tap impulse response sequence denoted by $\{\bar{\mathbf{H}}_{\mathrm{v}}[l]\in\mathbb{C}^{2J\times 2J}|\forall l\in\{1,\ldots,L\}\}$ through discrete Fourier transform, expressed as $\mathbf{H}_{\mathrm{v},k}=\sum_{l=1}^L\bar{\mathbf{H}}_{\mathrm{v}}[l]\mathrm{exp}(-j2\pi(k-1)(l-1)/K)$, where $\bar{\mathbf{H}}_{\mathrm{v}}[l]$ indicates the impulse response for the virtual channel at the $l$th tap which are assumed to be independent among the $L$ taps. We assume a rich scattering environment with Rayleigh fading so that each element in $\bar{\mathbf{H}}_{\mathrm{v}}[l]$ follows i.i.d. $\mathcal{CN}(0,P_l)$, where $P_l$ is the power of the $l$th-tap channel. $\mathbf{H}_{\mathrm{bs},k} \triangleq \mathbf{F}_{\mathrm{bs}}^T\mathbf{H}_{\mathrm{v},k}\mathbf{E}_\mathrm{bs}$ is defined as the equivalent beamspace MIMO channel at the $k$th subcarrier with each entry following i.i.d. $\mathcal{CN}(0,P)$, where $P=\sum_{l=1}^LP_l$, due to the orthonormality of $\mathbf{E}_\mathrm{bs}$ and $\mathbf{F}_\mathrm{bs}$.

\vspace{-0.3cm}
\subsection{Beamspace OFDM Symbol Transmission Using ESPAR}
After OFDM modulation at the transmitter, we can obtain the time-domain OFDM symbol $\{\mathbf{x}_\mathrm{bs}[t]\in\mathbb{C}^{N\times1}|t=\{1,\ldots,K,K+1,\dots,K+v\}\}$ as being circularly convoluted with the impulse response of the equivalent beamspace MIMO channel, where $\mathbf{x}_\mathrm{bs}[t]$ represents the signal transmitted at the $t$th discrete time index. Therefore, to radiate $\mathbf{x}_\mathrm{bs}[t]$ to the wireless environment, the required radiation pattern excited by the ESPAR is given by  
\vspace{-0.1cm}
\begin{equation}
	\mathbf{e}[t]=\mathbf{E}_{\mathrm{bs}}\mathbf{x}_\mathrm{bs}[t]. 
	\label{symbol_e}
	\vspace{-0.1cm}
\end{equation} 
This can be achieved by tuning the variable reactive loads of ESPAR, which are collected into a diagonal matrix $\mathbf{X}^L=\mathrm{diag}(0,x_2^L,\ldots,x_M^L)$ with $x_m^L$ for $m=2,\ldots,M$ being the load reactance connected to the $m$th antenna. With a given load configuration $\mathbf{X}^L$, the radiation pattern excited by the ESPAR containing the $\theta$ and $\phi$ polarization components sampled over $J$ angles, denoted as $\mathbf{e}\left(\mathbf{X}^L\right)\in\mathbb{C}^{2J\times1}$, can be obtained as
\begin{equation}
	\mathbf{e}\left(\mathbf{X}^L\right) = \mathbf{E}_\mathrm{oc}\left(\mathbf{Z}+j\mathbf{X}^L\right)^{-1}\mathbf{v}_s,
	\label{ESPAR_e}
\end{equation}
where $\mathbf{v}_s=[v_s,0,\ldots,0]^T\in\mathbb{C}^{M\times1}$ with $v_s$ indicating the source voltage at the active antenna 1. $\mathbf{Z}\in\mathbb{C}^{M\times M}$ refers to the ESPAR impedance matrix with its diagonal entries representing the self-impedance of the antennas and non-zero off-diagonal elements denoting the coupling or trans-impedance among the antennas. $\mathbf{E}_\mathrm{oc}=[\mathbf{e}_1^\mathrm{oc},\ldots,\mathbf{e}_M^\mathrm{oc}]\in\mathbb{C}^{2J\times M}$ is the open-circuit radiation pattern matrix of the $M$-element ESPAR, where $\mathbf{e}_m^\mathrm{oc}\in\mathbb{C}^{2J\times1}$ for $m=1,\ldots,M$ indicates the radiation pattern of the $m$th antenna excited by a unit current while all the other antennas are open-circuited, containing the dual polarization components sampled over $J$ angles. 

Therefore, by optimizing the variable reactive loads $\mathbf{X}^L$ of ESPAR, its radiation pattern $\mathbf{e}(\mathbf{X}^L)$ (\ref{ESPAR_e}) can be adjusted to approximate $\mathbf{e}[t]=\mathbf{E}_{\mathrm{bs}}\mathbf{x}_\mathrm{bs}[t]$ (\ref{symbol_e}) for transmitting the signal $\mathbf{x}_\mathrm{bs}[t]$. This is realized by maximizing their absolute correlation coefficient to be as close to unity as possible. This optimization problem can be expressed as \cite{HAN_OFDM,PCDM}
\begin{equation}
	\mathbf{X}^L_o = \underset{x_L\leq x_m^L\leq x_H,\forall m}{\mathrm{argmax}}\; \frac{|\langle \mathbf{e}[t],\mathbf{e}(\mathbf{X}^L) \rangle|}{\left\lVert\mathbf{e}[t]\right\rVert\,\left\lVert\mathbf{e}(\mathbf{X}^L)\right\rVert},
	\label{loads_opti}
\end{equation}
where $x_L$ and $x_H$ are the lower and upper bound for each load reactance, respectively. The corresponding signal transmitted in beamspace by the ESPAR in practice is then given by
\begin{equation}
	\mathbf{x}_\mathrm{bs}^\mathrm{ESPAR}[t] = \mathbf{E}_\mathrm{bs}^H\mathbf{e}(\mathbf{X}^L_o)=\mathbf{E}_\mathrm{bs}^H\mathbf{E}_\mathrm{oc}\left(\mathbf{Z}+j\mathbf{X}^L_o\right)^{-1}\mathbf{v}_s,
	\label{ESPAR_symbol}
\end{equation}
which is expected to approach $\mathbf{x}_\mathrm{bs}[t]$. 

To reduce computational complexity and enhance feasibility, the codebook-based approach proposed in \cite{HAN_OFDM} is used to implement (5). In that approach, the time-domain OFDM symbols are quantized into $2^Q$ codewords for constructing a codebook, each of which is associated with an ESPAR reactive load configuration found from (\ref{loads_opti}), where $Q$ refers to the number of bits of the quantizer.          

\vspace{-0.1cm}
\section{Proposed Detection Technique}
\subsection{Beamspace Error Floor Analysis}
There is usually a difference between the required radiation pattern $\mathbf{e}[t]$ (\ref{symbol_e}) and that excited by ESPAR in practice $\mathbf{e}(\mathbf{X}^L)$ (\ref{ESPAR_e}) due to the absolute correlation coefficient between them being less than unity. This produces variation between the ideal signal to be transmitted $\mathbf{x}_\mathrm{bs}[t]$ and the signal transmitted by ESPAR in practice $\mathbf{x}^{\mathrm{ESPAR}}_\mathrm{bs}[t]$ (\ref{ESPAR_symbol}), and we refer to this as the \textquotedblleft model error\textquotedblright. Such error will even be larger when the codebook-based approach in \cite{HAN_OFDM} is leveraged because of the OFDM symbol quantization. To quantify such an error, we denote $\mathbf{s}_{k}^o\in\mathbb{C}^{N\times1}$, $\forall k\in\mathcal{K}$, as the ideal transmitted signal at the $k$th subcarrier, denote $\mathbf{s}_k^\mathrm{ESPAR}\in\mathbb{C}^{N\times1}$ as the corresponding signal transmitted by ESPAR in practice (frequency response of $\mathbf{x}^\mathrm{ESPAR}_\mathrm{bs}[t]$), and define $\mathbf{n}_{\mathrm{bs},k}=\mathbf{s}_k^\mathrm{ESPAR}-\mathbf{s}_k^o$ as the model error at the $k$th subcarrier. To investigate the influence of the model error on the BER performance of the ESPAR based single-RF MIMO-OFDM system, we rewrite the received signal at the $k$th subcarrier (\ref{system_model_freq}) as 
\vspace{-0.1cm}
\begin{equation}
	\mathbf{y}_k = \mathbf{H}_{\mathrm{bs},k}\mathbf{s}_k^\mathrm{ESPAR}+\mathbf{n}_k =  \mathbf{H}_{\mathrm{bs},k}\mathbf{s}^o_k+\mathbf{H}_{\mathrm{bs},k}\mathbf{n}_{\mathrm{bs},k}+\mathbf{n}_k.
	\label{yk_espar}
	\vspace{-0.1cm}
\end{equation}
It can be observed from (\ref{yk_espar}) that when using ESPAR to transmit OFDM symbols, the model error yields an additional noise term with power $\left\lVert\mathbf{H}_{\mathrm{bs},k}\mathbf{n}_{\mathrm{bs},k}\right\rVert^2$ which is non-white, independent of AWGN, and proportional to the transmit power. Therefore, increasing the transmit power cannot mitigate the effects of such a noise caused by the model error $\mathbf{n}_{\mathrm{bs},k}$, which can potentially influence the accuracy of the detected signals and yield irreducible BER at high transmit power, leading to an error floor. 

If maximum likelihood (ML) detection utilizing the conventional Euclidean distance is applied at the receiver, when the transmit power is high enough so that the effects of AWGN $\mathbf{n}_k$ can be neglected, the detected transmitted signal at the $k$th subcarrier can be expressed as
\vspace{-0.05cm}
\begin{equation}
	\begin{aligned}
	\hat{\mathbf{s}}_k^\mathrm{ML} &= \underset{\mathbf{s}_k\in\mathcal{S}}{\mathrm{argmin}}\; \left\lVert \mathbf{y}_k - \mathbf{H}_{\mathrm{bs},k}\mathbf{s}_k \right\rVert^2 \\
	& = \underset{\mathbf{s}_k\in\mathcal{S}}{\mathrm{argmin}}\; \left\lVert \mathbf{H}_{\mathrm{bs},k}\left(\mathbf{s}^o_k+\mathbf{n}_{\mathrm{bs},k}\right) - \mathbf{H}_{\mathrm{bs},k}\mathbf{s}_k \right\rVert^2,
	\label{ML}
	\end{aligned}
	\vspace{-0.05cm}
\end{equation}
where $\mathcal{S}$ is the set containing all the modulation constellations of the transmitted signals. It can be noticed from (\ref{ML}) that $\mathbf{n}_{\mathrm{bs},k}$ is associated with $\mathbf{H}_{\mathrm{bs},k}$. This implies that the model error can be significant and the stochastic distribution characteristics can be potentially varied because of the nonuniform singular values of the MIMO channel matrix. The BER performance degradation can thus be significant and sensitive to even small model errors when using conventional ML detection due to the influence from the channel, yielding irreducible BER at high transmit power. On the other hand, if minimum mean square error (MMSE) detection is used, the detected signal at high transmitted power is given by
\vspace{-0.1cm}
\begin{equation}
	\hat{\mathbf{s}}_k^\mathrm{MMSE} = \left(\mathbf{H}_{\mathrm{bs},k}^H\mathbf{H}_{\mathrm{bs},k}\right)^{-1}\mathbf{H}_{\mathrm{bs},k}^H\mathbf{y}_k = \mathbf{s}_k^o+\mathbf{n}_{\mathrm{bs},k}.
	\vspace{-0.1cm}
\end{equation}
It can be noted that the model error is decoupled from the channel. As long as $\mathbf{n}_{\mathrm{bs},k}$ is not large enough to exceed the correct hard-decision boundary, i.e. each entry in $\mathbf{n}_{\mathrm{bs},k}$ has absolute value less than half of the distance between constellation points, the model error caused by using ESPAR for transmission has no influence on the BER performance of the single-RF MIMO-OFDM system. Therefore, it can be concluded that MMSE detection is more robust to the model error compared with the conventional ML detection utilizing the Euclidean distance. 

\vspace{-0.2cm}
\subsection{Covariance-Aware Detection}
Based on the error floor analysis and the observation that employing MMSE detection can potentially eliminate the effects of the model error, we propose the covariance-aware detection approach to handle the colored noise introduced by the model error of ESPAR \cite{ML1,ML2}. When using the Euclidean distance $\left\lVert \mathbf{y}_k - \mathbf{H}_{\mathrm{bs},k}\mathbf{s}_k \right\rVert^2$, it is implicitly assumed that the noise is i.i.d. isotropic white noise. However, the existence of the model error combined with the channel makes the entries in the total noise correlated, which has been ignored in the conventional ML detection implementing the Euclidean distance. To characterize this correlation and following the filter approach in MMSE detection, we use (\ref{yk_espar}) to define the covariance matrix of the total noise with given channel state information (CSI) at the $k$th subcarrier as 
\vspace{-0.05cm}
\begin{equation}
\begin{aligned}
\mathbf{R}_k &= \mathbb{E}\left\{\left(\mathbf{H}_{\mathrm{bs},k}\mathbf{n}_{\mathrm{bs},k}+\mathbf{n}_k\right)\left(\mathbf{H}_{\mathrm{bs},k}\mathbf{n}_{\mathrm{bs},k}+\mathbf{n}_k\right)^H\right\}  \\
&=\mathbf{H}_{\mathrm{bs},k}\mathbb{E}\left\{\mathbf{n}_{\mathrm{bs},k}\mathbf{n}_{\mathrm{bs},k}^{H}\right\}\mathbf{H}_{\mathrm{bs},k}^H + N_{0}\mathbf{U}_R,
\label{Rk}
\end{aligned}
\vspace{-0.05cm}
\end{equation}
where $\mathbb{E}\{\mathbf{n}_{\mathrm{bs},k}\mathbf{n}_{\mathrm{bs},k}^{H}\}$ is the model error covariance matrix. Its diagonal entries indicate the power of the model error on each of the $N$ orthogonal basis radiation patterns and its off-diagonal elements represent their correlations. To quantify the model error introduced by the ESPAR, we define its average power across pattern basis and subcarrier frequencies as  
\vspace{-0.1cm} 
\begin{equation}
N_\mathrm{bs}=\mathbb{E}\Biggl\{ \frac{1}{KN}\sum_{k=1}^K\bigl\lVert \underset{\mathbf{n}_{\mathrm{bs},k}}{\underbrace{\mathbf{s}_k^\mathrm{ESPAR}-\mathbf{s}_k^o}}\bigl\rVert^2 \Biggl\}.
\label{Nbs}
\vspace{-0.1cm}
\end{equation} 

Leveraging the covariance of the total noise (\ref{Rk}), our approach to detect the signal at the $k$th subcarrier is to use  
\vspace{-0.05cm} 
\begin{equation}
	\hat{\mathbf{s}}_k^\star = \underset{\mathbf{s}_k\in\mathcal{S}}{\mathrm{argmin}}\; \left( \mathbf{y}_k - \mathbf{H}_{\mathrm{bs},k}\mathbf{s}_k \right)^H \mathbf{R}_k^{-1}\left( \mathbf{y}_k - \mathbf{H}_{\mathrm{bs},k}\mathbf{s}_k \right),
	\label{detector_proposed}
	\vspace{-0.05cm}
\end{equation}
which is equivalent to minimizing the squared Mahalanobis distance \cite{Madis1}. This is inspired by evaluating the Euclidean distance after the colored Gaussian noise is whitened \cite{ML1,ML2} as in a conventional ML detector. While we have not proven the total noise is Gaussian distributed, our simulation results in the following section show the formulation is effective.

The implementation of $\mathbf{R}_k^{-1}$ in (\ref{detector_proposed}) can be interpreted as decorrelating and whitening the colored noise through transforming the received signal space before the exhaustive search so that the noise can be decoupled from the channel, analogously to MMSE detection. It can also be understood as adapting the decision region according to the distorted noise distribution. Thus, this covariance-aware detection can handle the error floor and thus further enhance the BER performance.

\section{Simulation Results}
\subsection{ESPAR Design and Orthogonal Basis Radiation Patterns}
\begin{figure}[t]
	\centering
	\includegraphics[scale=0.54]{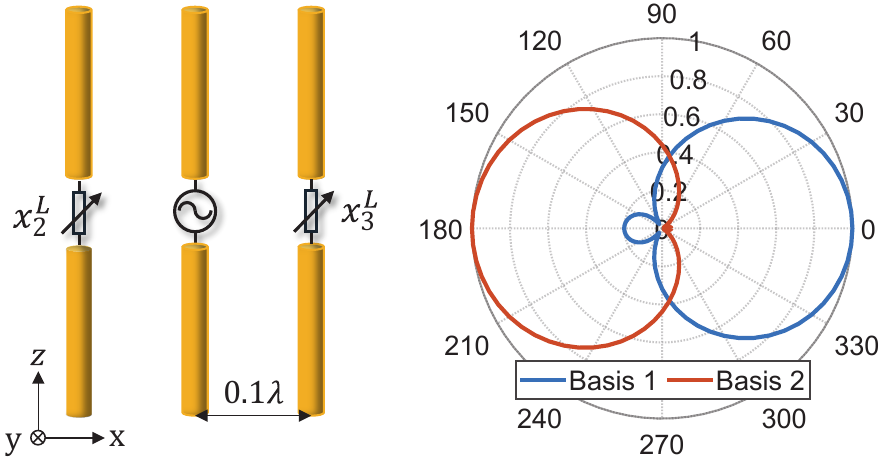}
	\vspace{-0.3cm}
	\caption{ESPAR consisting of 3-element linear dipole array with center frequency 2.45 GHz (left) and its orthogonal basis radiation patterns (right) for transmitting beamspace symbols in the MIMO-OFDM system for 2D-PAS. }
	\vspace{-0.3cm}
	\label{Fig_ESPAR}
\end{figure}

To evaluate the performance of the proposed detection technique for solving the error floor problem in single-RF MIMO-OFDM systems using ESPAR, a 3-element linear dipole array is considered at the transmitter as shown in Fig. \ref{Fig_ESPAR}. An active dipole antenna is at the center and connected to a single-RF source, having length 47 mm and radius 2 mm. Another two identical parasitic dipole antennas are placed on either side of the center dipole with a spacing of 0.1 wavelength ($\lambda$) and connected to reactive loads. Considering the feasibility, we set the lower and upper bound on the load reactance as $x_L=-100$ Ohm and $x_H=100$ Ohm, respectively \cite{HAN_OFDM}. The impedance matrix $\mathbf{Z}$ and the open-circuit radiation patterns matrix $\mathbf{E}_\mathrm{oc}$ in (\ref{ESPAR_e}) of the ESPAR are obtained by CST studio suite through full-wave electromagnetic simulation \cite{CST}.    

We consider two propagation environments both with a cross polarization discrimination of unity. The first is with a 2-D uniform power angular spectrum (PAS) over the azimuth angle on the XOY plane, denoted as 2D-PAS. The second is a full 3-D uniform PAS which we denote as 3D-PAS. Under both scattering environments with the ESPAR as presented in Fig. \ref{Fig_ESPAR}, it is found that by applying the pattern correlation decomposition method \cite{PCDM}, the EADoF of the system is 2 and thus $N=2$ individual OFDM symbol streams can be transmitted simultaneously in beamspace. The reactive loads of ESPAR are optimized by the interior-point method to generate the two orthogonal basis radiation patterns. For the 2D-PAS environment, these on the azimuth plane are shown in Fig. \ref{Fig_ESPAR}. The corresponding reactive loads of the two parasitic dipoles for forming Basis 1 and Basis 2 are $x_2^L=2.3$ nH, $x_3^L=5.0$ pF and $x_2^L=1.5$ pF, $x_3^L=2.8$ nH, respectively. 

\vspace{-0.1cm}
\subsection{BER Performance Evaluation for QPSK and 2D-PAS}
Given $N=2$ we consider a 2$\times$2 beamspace MIMO-OFDM system with $K=64$ subcarriers using QPSK modulation. The ESPAR shown in Fig. \ref{Fig_ESPAR} is used at the transmitter and two ideally isolated conventional antennas synchronized with the transmitter are at the receiver under a 2D-PAS environment. We assume that the frequency selective channel has $L=8$ taps with power of $[0,-2,\dots,-14]$ dB and the total power of the channel is normalized to unity, i.e. $P=\sum_{l=1}^LP_l = 1$. The length of CP is set to $v=16$, quarter of the number of subcarriers. 

To transmit the OFDM symbols using ESPAR, we adopt the codebook-based approach as proposed in \cite{HAN_OFDM} using 8-bit and 12-bit quantizers. The codebooks construction and the estimation of the model error covariance matrix at each subcarrier are carried out offline before transmission and these are shared with the transmitter and receiver synchronously. To estimate the model error covariance matrix in (\ref{Rk}), $10^6$ simulated frequency-domain signals are randomly generated at each subcarrier and transmitted by the ESPAR using the method proposed in \cite{HAN_OFDM} after OFDM modulation. $\mathbb{E}\{\mathbf{n}_{\mathrm{bs},k}\mathbf{n}^H_{\mathrm{bs},k}\}$ can thus be approximated by averaging over these $10^6$ realizations of $\mathbf{n}_{\mathrm{bs},k}$, $\forall k$. The average powers of the model errors as defined in (\ref{Nbs}), $N_{\mathrm{bs}}$, for the 8-bit and 12-bit quantizers are 0.016 and 0.007, respectively. 

\subsubsection{Perfect CSI}
\begin{figure}[t]
	\centering
	\includegraphics[scale=0.31]{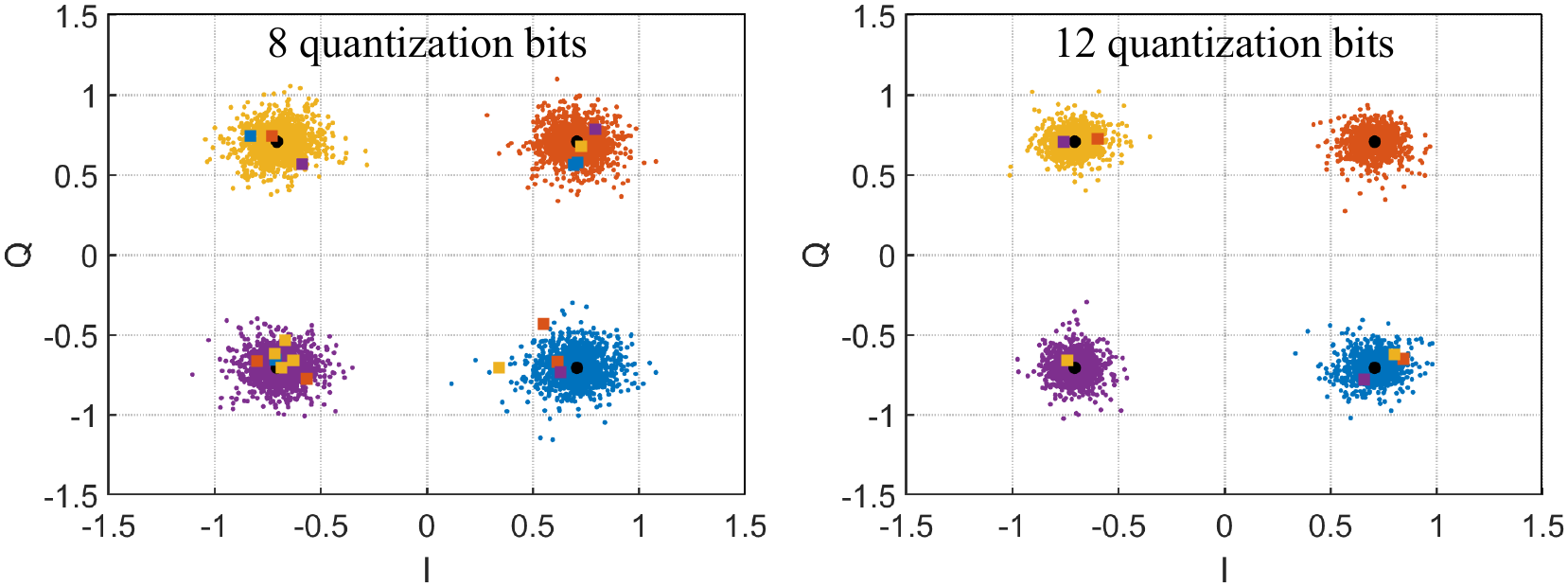}
	\vspace{-0.3cm}
	\caption{Constellations of the single-RF MIMO-OFDM system using ESPAR with 8-bit (left) and 12-bit (right) quantizers at $E_\mathrm{b}/N_0 = 50$ dB. Black dots represent ideal QPSK signals.}
	\vspace{-0.5cm}
	\label{constellation}
\end{figure}

We first simulate the BER of the MIMO-OFDM system under perfect CSI by performing conventional MMSE, ML, and our proposed detection approaches, as shown in Fig. \ref{Fig_BER}. The BER of the conventional MIMO-OFDM system is also provided as a benchmark, where two ideally isolated conventional antennas are used at the transmitter. To provide more insights and for detailed demonstration, the constellations at $E_\mathrm{b}/N_0=50$ dB are presented in Fig. \ref{constellation}, where $E_\mathrm{b}$ refers to the energy per information bit which is set to be high so that the influence of AWGN can be neglected. Specifically, the dots plotted in Fig. \ref{constellation} indicate $\mathbf{H}_{\mathrm{bs},k}^{-1}\mathbf{y}_k$ and their distance from the ideal QPSK signals are attributed to the model error introduced by ESPAR from (\ref{yk_espar}). The dots and squares with identical color are detected as the same QPSK signal by the ML detection using Euclidean distance. The squares indicate the received signals located at the correct quadrature but being incorrectly detected by the conventional ML detector, in which case performing MMSE detection yields a BER of zero.

From the results presented in Fig. \ref{Fig_BER} and Fig. \ref{constellation}, four observations can be made. \textit{Firstly}, the degradation of the BER by performing MMSE detection for the beamspace MIMO-OFDM system is negligible and no error floor occurs at high transmit power. This implies that the MMSE detection is insensitive and has high robustness to the model error, demonstrating the rationality of our proposed detection method leveraging the characteristics of the MMSE detection approach. \textit{Secondly}, the conventional ML detector is sensitive to the model error introduced by using ESPAR to transmit OFDM symbols and an obvious error floor occurs at high transmit power, with performance worse than MMSE. \textit{Thirdly}, when the number of quantization bits $Q$ increases, it can be noticed from Fig. \ref{constellation} that the received signals are more concentrated around the ideal QPSK signals and those incorrectly detected by the conventional ML detection technique are fewer, implying less model error being introduced by ESPAR, i.e. $N_\mathrm{bs}\propto Q^{-1}$. This is also revealed in Fig. \ref{Fig_BER} that when more quantized codewords are utilized for transmission, the BER performance is improved and the error floor tends to be alleviated but still exists. \textit{Lastly}, compared to the BER by applying conventional ML detection, leveraging our detector can significantly alleviate the error floor and further enhance the BER performance of the single-RF MIMO-OFDM system as revealed in Fig. \ref{Fig_BER}. For illustration, approximately 5 dB energy per bit can be saved when $Q=8$ to achieve a BER of $10^{-3}$ by using our detector.  

\subsubsection{Imperfect CSI}

\begin{figure}[t]
	\centering
	\includegraphics[scale=0.39]{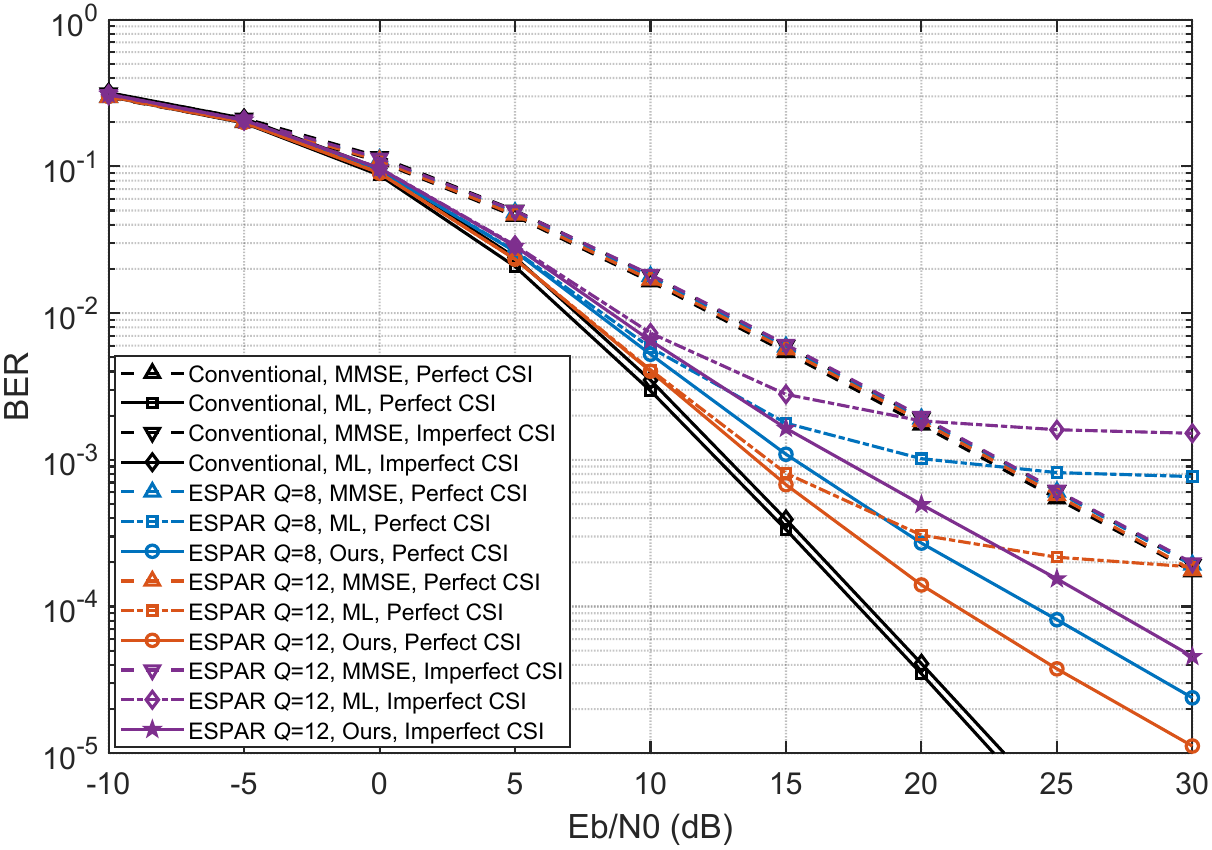}
	\vspace{-0.3cm}
	\caption{BER performance of the single-RF MIMO-OFDM system using ESPAR versus $E_\mathrm{b}/N_0$ under perfect and imperfect CSI with QPSK modulation and 2D-PAS scattering environment. }
	\vspace{-0.2cm}
	\label{Fig_BER}
\end{figure}

We next evaluate the BER performance of our proposed detection technique under imperfect CSI since from (\ref{Rk}), it also depends on the accuracies of the estimated channel matrix apart from the model error covariance matrix. The results using the 12-bit quantizer are presented in Fig. \ref{Fig_BER}. The block-type pilots are used for channel estimation and 14 OFDM symbols are assumed per symbol block with the [2, 5, 10, 13]th symbols used as pilots. The conventional MMSE channel estimator is applied. From Fig. \ref{Fig_BER}, it can be observed that when applying our detector and the conventional ML detector, the BER of the ESPAR enabled system is more sensitive to the channel estimation errors, compared with the conventional system. This can be attributed to the pilot variations caused by the model error when using ESPAR for transmission, which can then influence the accuracy of the channel estimation and thus the evaluation of the required Mahalanobis distance during detection. Despite such a BER performance degradation, results demonstrate that the error floor can still be effectively alleviated by our proposed detection method.

\vspace{-0.2cm}
\subsection{BER Performance Evaluation with 16-QAM and 3D-PAS}
\begin{figure}[t]
	\centering
	\includegraphics[scale=0.39]{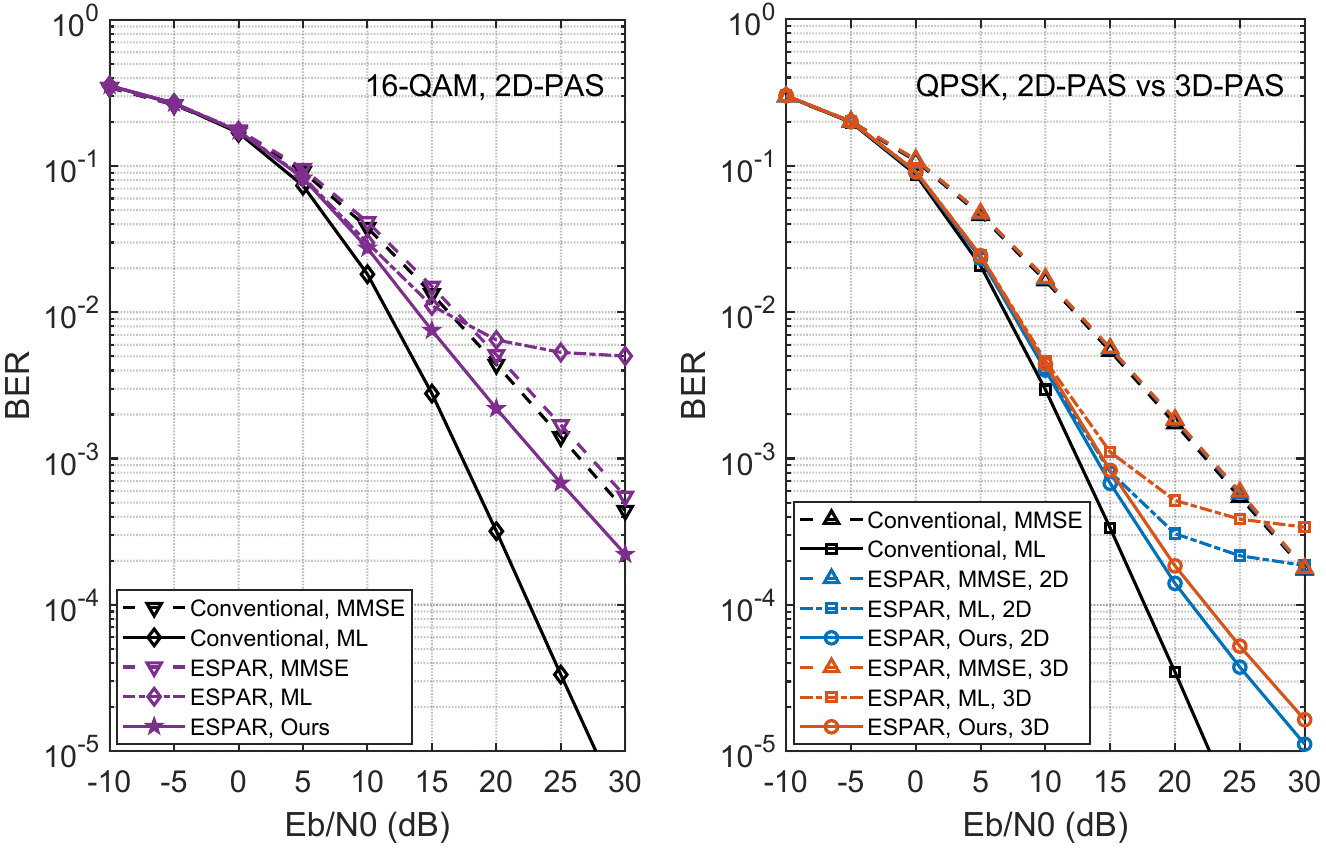}
	\vspace{-0.7cm}
	\caption{BER performance of the ESPAR enabled single-RF MIMO-OFDM system with $Q=12$ under 16-QAM (left) and 3D-PAS (right). }
	\vspace{-0.5cm}
	\label{Fig_BER2}
\end{figure}

To provide more insights, we next evaluate the BER performance of our proposed detector under more complicated scenarios assuming perfect CSI. We first explore the effects of higher-order modulation by considering 16-QAM. From the simulated BER results in Fig. \ref{Fig_BER2}, it can be found that the BER gap between our detector and the conventional system becomes larger while the error floor is still effectively addressed. For 16-QAM modulation, the power of the model error is $N_{\mathrm{bs}}=0.007$ which is identical to the scenario using QPSK modulation. This is because when constructing the codebook as in \cite{HAN_OFDM}, the utilized training sequence is generated by normalizing the OFDM symbols whose entries satisfy a Gaussian distribution so that the codebook can largely characterize OFDM symbols with different modulation types. The reason why the conventional ML detection suffers more seriously from the model error and the BER performance improvement of our proposed detector is less obvious is that when $E_{\mathrm{b}}/N_{0}$ is the same, the received signals are more easily detected incorrectly since the distances among constellation points are closer when the modulation order is higher. 

We next investigate the influence of the scattering scenario by consisting the 3D-PAS environment. The BER results for QPSK are shown in Fig. \ref{Fig_BER2} and $N_{\mathrm{bs}}$ is 0.01 (for $Q=12$). The $N_{\mathrm{bs}}$ is larger than for 2D-PAS because when optimizing the correlation between radiation patterns excited by the ESPAR and the desired ones as described in (\ref{loads_opti}), the patterns over the full 3-D sphere need to be considered, while for the 2-D case, only the patterns on a plane are required to be designed. Therefore, the patterns radiated by the ESPAR can approach the desired ones more closely for the more limited scattering 2-D case, yielding smaller model error. As a result, with identical QPSK modulation, the BER performance of the ESPAR enabled system is better when the main PAS is concentrated within a smaller spatial angle region.      

From the presented results, it can be concluded that when the total noise covariance matrix is considered for detection, the error floor problem that exists in the single-RF system can be effectively alleviated. The BER performance of our proposed detection technique, i.e. its gap with the BER of the conventional systems, is related to the accuracies of the estimation for the total noise covariance matrix. It also depends on the relative magnitude of the model error compared with the distance among constellation points, which can be influenced by the quantization level, order of modulation, and the scattering characteristics. 

\vspace{-0.3cm}  
\subsection{Computational Complexity Analysis}
The BER performance gains of our proposed detection approach are achieved at the expense of increased computational complexity due to the additional covariance matrix introduced in the evaluation of the Mahalanobis distance, as revealed in (\ref{detector_proposed}). To investigate such a tradeoff, we quantify the computational complexity per subcarrier for the conventional MMSE, ML, and our proposed detectors as summarized in Table I, where $J$ indicates the order of modulation. It should be noted that for our detector, the model error covariance matrix is obtained in advance and $\mathbf{R}_k$ with its inverse is only required to be calculated once before the exhaustive search at each subcarrier. The $E_{\mathrm{b}}/N_{0}$ comparisons of the conventional ML and our proposed detectors for achieving identical BER levels under perfect CSI, QPSK modulation, and 2D-PAS environment are also compared in Table I, benchmarked with the MMSE detector. From Table I, it can be found that by using our proposed detection technique, the $E_{\mathrm{b}}/N_{0}$ required to realize the targeted BER can be usefully reduced, compared with the conventional MMSE and ML detectors. Although such a significant improvement brings higher computational costs, the increase in computational complexity occurs at the receiver in the access point or base station and these are typically not as restricted by computing resources as the users or sensor devices.   

\begin{table}[t]
	\caption{Computational Complexity and BER Performance Comparison with the MMSE Detector as Baseline}
	\vspace{-0.2cm}
	\centering{}%
	\begin{tabular}{>{\centering}m{0.8cm}>{\centering}m{2.4cm}>{\centering}m{1.86cm}>{\centering}m{1.86cm}}
		\toprule 
		\multirow{2}{0.8cm}{\centering{}Method\vspace{-0.3cm}} & \multirow{2}{2.4cm}{\centering{}Computational Complexity\vspace{-0.3cm}} & \multicolumn{2}{c}{$E_{\mathrm{b}}/N_{0}$ Improvement\vspace{-0.1cm}}\tabularnewline\addlinespace[0.1cm]
		\cmidrule{3-4}
		&  & \centering{}@ BER$=10^{-3}$ & \centering{}@ BER$=10^{-4}$\tabularnewline
		\midrule
		\midrule 
		MMSE & $\mathcal{O}(N^{2}(N+R))$ & \multicolumn{2}{c}{0 dB (Benchmark)}\tabularnewline
		\midrule 
		ML & $\mathcal{O}(J^{N}RN)$ & 3 dB & Error Floor\tabularnewline
		\midrule 
		Ours & $\mathcal{O}(J^{N}R(N+R))$ & 8 dB 
		
		($Q=8$) & 11 dB ($Q=12$)\tabularnewline
		\bottomrule
	\end{tabular}
	\vspace{-0.3cm}
\end{table}

\vspace{-0.2cm}
\subsection{Further Enhancements}
To further enhance the performance, quantizers with larger numbers of quantization bits, more advanced channel estimation techniques such as the orthogonal matching pursuit recovery algorithm \cite{ChannelEstimation}, and online calibration of the model error covariance matrix can be implemented. Other advanced detection techniques can also be investigated in the application scenario of single-RF MIMO systems to further improve the robustness against the estimation inaccuracies while also reducing the computational complexity of the receiver, especially for higher-order modulation and larger-scale MIMO systems. In addition, more advanced reconfigurable antennas such as the highly reconfigurable pixel antenna with a single feeding port can be exploited to replace the ESPAR at the transmitter for forming the single-RF system \cite{AntennaCoding1,AntennaCoding2}. Compared to ESPAR, the pixel antenna typically has dozens more design variables that can be optimized and thus can be flexibly adjusted providing higher pattern accuracy. Therefore, it is expected that the pixel antenna can better approximate the ideal transmitted signals and have less model error, leading to improved BER performance.
 
\section{Conclusion}
In this work, the error floor in single-RF MIMO-OFDM systems at high transmit power has been systematically analyzed. Such an error floor is attributed to the additional model error introduced by using reconfigurable antennas to transmit OFDM symbols. To address the error floor and further improve the BER performance, the application of a covariance-aware detection technique has been proposed. Simulation results taking ESPAR as an example of a reconfigurable antenna under various communication scenarios demonstrate the effectiveness and generality of our proposed approach.

\end{document}